\documentclass[12pt]{iopart}
\usepackage{graphicx} 

\begin{document}

\title[Electron transport through strongly interacting quantum dot]
{Electron transport through strongly interacting quantum dot coupled 
 to normal metal and superconductor.}

\author{Mariusz Krawiec\dag\ and Karol I. Wysoki\'{n}ski\ddag}

\address{\dag\ H. H. Wills Physics Laboratory, 
 University of Bristol, Tyndall Ave., Bristol BS8 1TL, UK}

\address{\ddag\ Institute of Physics, M. Curie-Sk\l odowska University,
ul. Radziszewskiego 10a, 20-031 Lublin, Poland}
 
\begin{abstract}
We study the electron transport through the quantum dot coupled to the normal
metal and $BCS$-like superconductor ($N - QD - S$) in the presence of the Kondo 
effect and Andreev scattering. The system is described by the single impurity 
Anderson model in the limit of strong on-dot interaction. We use recently 
proposed equation of motion technique for Keldysh nonequilibrium Green's 
function together with the modified slave boson approach to study the electron 
transport. We derive formula for the current which contains various tunneling 
processes and apply it to study the transport through the system. We find that 
the Andreev conductance is strongly suppressed and there is no zero-bias 
(Kondo) anomaly in the differential conductance. We discuss effects of the 
particle-hole asymmetry in the electrodes as well as the asymmetry in the
couplings.
\end{abstract}
\pacs{74.50.+r, 72.15.Qm, 73.23.Hk}

\maketitle


\section{\label{sec1} Introduction}

In recent years there has been much experimental and theoretical work on
electron transport through nanometer-size areas  (metallic or semiconducting 
islands) containing small number of electrons. These islands (sometimes called 
the quantum dots) are coupled via tunnel barriers to several external 
electrodes making it possible to adjust the current flowing through the system 
\cite{Ferry}. The devices give a new possibility of studying several well-known 
quantum phenomena in novel and highly controllable way. For instance, it is 
well known, that quantum dot behaves like magnetic impurity in a metallic host 
and in particular displays the Kondo effect \cite{Glazman}-\cite{Hershfield}.
Kondo effect is a manifestation of the simplest state formed by the impurity
spin and conduction electron spins. This state gives rise to a quasiparticle 
peak at the Fermi energy in the dot spectral function 
\cite{MeirWingreen}-\cite{HettlerSchoeller} and zero-bias maximum in the 
differential conductance observed experimentally 
\cite{GoldhaberGordon}-\cite{Sasaki}.

Another example is the Andreev scattering \cite{Andreev}, according to which an 
electron impinging on normal metal - superconductor interface is reflected back 
as a hole and the Cooper pair is created in superconductor. This effect has 
been shown to play crucial role in the transport properties of various hybrid 
mesoscopic superconducting devices \cite{LambertRaimondi}. There is a number of 
papers in the literature concerning the electron transport in various 
realizations of such devices. Here we are interested in study of the normal 
state quantum dot coupled to one normal and one superconducting electrode 
($N - QD - S$). Such system was studied within scattering matrix technique 
\cite{Beenakker, Lambert}. However this approach is valid only for 
noninteracting systems and cannot take into account effects of Coulomb 
interactions between electrons on the dot, which are very important in these 
small systems, as they  lead e.g. to the Coulomb blockade phenomena 
\cite{Kouwenhoven} or Kondo effect \cite{Glazman}-\cite{Hershfield}. Transport 
through noninteracting quantum dot has also been studied within nonequilibrium 
Green's function technique. The effect of multiple discrete levels of the dot 
was discussed in Refs.\cite{SunWang,FengXiong}, the photon-assistant transport 
in \cite{ZhaoGehlen}, electron transport in the weak magnetic field in 
\cite{Zhao}, temperature dependence of the resonant Andreev reflections in 
\cite{Zhutemp} and transport in three terminal system (two ferromagnetic and 
one superconducting electrodes) in \cite{ZhuFM}.

In the presence of strong Coulomb interaction in $N - QD - N$ device the Kondo 
effect appears an influences the  electron transport in the system. If one of
the electrodes is superconducting both single electron and the Andreev current
is affected by the Abrikosov - Suhl resonance. This problem has been 
extensively studied within various techniques 
\cite{FengXiong,Fazio}-\cite{SunGuo} and there is no consensus. Some authors 
have predicted suppression \cite{FengXiong,Fazio,Clerk,Avishai} of the 
conductance due to Andreev reflections while others - enhancement 
\cite{Schwab,Kang}. Recently it has been shown \cite{Cuevas} that one can 
obtain either suppression or enhancement of the conductance in dependence on 
the values of the model parameters. Recently other effects like emergence of 
the Kondo-like peaks in the local density of states ($DOS$) at energies equal 
to $\pm \Delta$ ($\Delta$ is the superconducting order parameter) 
\cite{Clerk,Cuevas} or a novel co-tunneling process (Andreev-normal 
co-tunneling) \cite{SunGuo} have been revealed. This process involves Andreev 
tunneling from the $QD - S$ interface and normal tunneling from $N - QD$ 
interface. As a result the Cooper pair directly participates in the formation 
of the spin singlet (Kondo effect) and leads to emergence of the additional 
Kondo resonances in the local $DOS$ and enhancement of the tunneling current.

The purpose of the present work is to apply the new technique to derive formula 
for the current through $QD$ (in the limit of strong on-dot Coulomb 
interaction) in terms of various tunneling processes. We also study the 
interplay between the Kondo effect and Andreev reflections to give additional 
insight into the the problem of the suppression/enhancement of the zero-bias 
current-voltage anomaly. Further we discuss the question of participation of 
the superconducting electrons in creation of the Kondo effect. And finally we
investigate the influence of the electron - hole asymmetry in the leads on
tunneling transport as well as the asymmetry in the couplings to the leads.

The paper is organized as follows. In section \ref{sec2} we present model under
consideration and derive formula for the current using $EOM$ for nonequilibrium
$GF$. In section \ref{sec3} we apply the obtained formula for the current to 
the numerical study of the transport through $N - QD - S$ system. Section 
\ref{sec4} is devoted to the interplay between the Kondo and Andreev 
scattering. In section \ref{sec5} we discuss the quantum dot asymmetrically 
coupled to the leads, while the effect of electron - hole asymmetry in the 
leads is investigated in sec. \ref{sec6}. Some conclusions are given in sec. 
\ref{sec7}.


\section{\label{sec2} Model and formulation}

The Anderson Hamiltonian of the single impurity \cite{Anderson}, in the Nambu
representation, can be written in the form
\begin{eqnarray}
H = \sum_{\lambda {\bf k} \sigma} \Psi^+_{\lambda {\bf k} \sigma} 
H^0_{\lambda {\bf k}} \Psi_{\lambda {\bf k} \sigma} +
\nonumber \\
\sum_{\sigma} \Phi^+_{\sigma} H^{QD} \Phi_{\sigma} +
\sum_{\lambda {\bf k} \sigma} \Psi^+_{\lambda {\bf k} \sigma}
H^I_{\lambda {\bf k}} \Phi_{\sigma}
\label{HAND}
\end{eqnarray}
where the Nambu spinors $\Psi_{\lambda {\bf k} \sigma}$ and $\Phi_{\sigma}$ are
defined as
\begin{eqnarray}
\Psi_{\lambda {\bf k} \sigma} = \left(
\begin{array}{l}
c_{\lambda {\bf k} \sigma} \\
c^+_{\lambda {-\bf k} -\sigma}
\end{array}
\right) 
\;\;\;\;\;\; ; \;\;\;\;\;\;
\Phi_{\sigma} = \left(
\begin{array}{l}
d_{\sigma} \\
d^+_{-\sigma}
\end{array}
\right)
\label{PSIPHI}
\end{eqnarray}
and
\begin{eqnarray}
H^0_{\lambda {\bf k}} = \left(
\begin{array}{cr}
\epsilon_{\lambda {\bf k}} & \Delta_{\lambda {\bf k}} \\
\Delta^*_{\lambda {\bf k}} & -\epsilon_{\lambda {\bf k}}
\end{array}
\right) 
\label{H0}
\end{eqnarray}
denotes Hamiltonian of the normal ($\Delta_{N {\bf k}} = 0$) or superconducting
($\Delta_{S {\bf k}} \neq 0$) lead.
\begin{eqnarray}
H^{QD} = \left(
\begin{array}{cc}
E_d + U_d n_{-\sigma} & 0 \\
0 & - ( E_d + U_d n_{\sigma})
\end{array}
\right) 
\label{HQD}
\end{eqnarray}
is the dot Hamiltonian and 
\begin{eqnarray}
H^I_{\lambda {\bf k}} = \left(
\begin{array}{cr}
V_{\lambda {\bf k}} & 0 \\
0 & -V_{\lambda {\bf k}}
\end{array}
\right) 
\label{HI}
\end{eqnarray}
the dot-electrode hybridization. Here $\lambda = N$, $S$ denotes the normal 
metal ($N$) or superconducting ($S$) lead in the system. The parameters have 
the following meaning: $c^+_{\lambda {\bf k} \sigma}$ 
($c_{\lambda {\bf k} \sigma}$) denotes creation (annihilation) operator for a 
conduction electron with the wave vector ${\bf k}$, spin $\sigma$ in the lead 
$\lambda$, $\Delta_{\lambda {\bf k}}$ is the superconducting order parameter 
in the lead $\lambda$ ($\Delta_{S {\bf k}} = \Delta_S$, 
$\Delta_{N {\bf k}} = 0$) , and $V_{\lambda {\bf k}}$ is the hybridization 
matrix element between conduction electron of energy 
$\epsilon_{\lambda {\bf k}}$ in the lead $\lambda$ and localized electron on 
the dot with the energy $E_d$. $d^+_{\sigma}$ ($d_{\sigma}$) is the creation 
(annihilation) operator for an electron on the dot and $U_d$ is the on-dot 
Coulomb repulsion.

To derive the formula for the average current in the system we start from the 
time derivative of the charge (for convenience we perform calculations in the 
normal electrode) \cite{Yauho}:
\begin{eqnarray}
J = - e \frac{d}{d t} \langle N_N \rangle= 
\frac{ie}{\hbar} \langle [N_N, H]_- \rangle 
\label{CURDEF}
\end{eqnarray},
where $N_N = \sum_{{\bf k} \sigma} c^+_{N {\bf k} \sigma} c_{{\bf k} \sigma}$ 
is the total electron number operator in the lead $N$ and $e$ is the elementary
charge. The above formula can be written in terms of the Green's functions
($GF$):
\begin{eqnarray}
J = - \frac{2e}{\hbar} \sum_{{\bf k} \sigma} 
\int^{\infty}_{-\infty} \frac{d\omega}{2\pi} 
[H^I_{N {\bf k}} G^<_{N {\bf k} \sigma, d}(\omega)]_{11}
\label{CURR1}
\end{eqnarray},
where $G^<_{N {\bf k}, d \sigma}(\omega)$ is the Fourier transform of the
Keldysh matrix Green's function \cite{Keldysh} 
$G^<_{N {\bf k}, d \sigma}(t) = 
i \langle \Phi^+_{\sigma}(0) \otimes \Psi_{N {\bf k} \sigma}(t) \rangle$.
Now we have to calculate the Green's function 
$G^<_{N {\bf k}, d \sigma}(\omega)$. One can do this in the usual way, i. e..
using Keldysh equation \cite{Keldysh,Yauho} 
\begin{eqnarray}
G^<(\omega) = (1 + G^r(\omega) \Sigma^r(\omega)) G^<_0(\omega) 
\nonumber \\
\times
(1 + \Sigma^a(\omega) G^a(\omega)) + 
G^r(\omega) \Sigma^<(\omega) G^a(\omega)
\label{KELDEQ}
\end{eqnarray}
(superscripts $r, a$ are for retarded and advanced $GF$
respectively) and make more or less justified approximations for the 'lesser' 
self-energy $\Sigma^<(\omega)$.  Usually one uses approximation due to Ng 
\cite{Ng} which  states that full 'lesser' self-energy is proportional to the 
noninteracting one ($\Sigma^<(\omega) \propto \Sigma^<_0(\omega)$). This 
approximation is widely used in the literature \cite{Fazio,Raimondi}. However 
we wish to use another approach based on recently proposed equation of motion 
technique ($EOM$) for nonequilibrium Green's functions \cite{Niu}. The usual 
equation of motion derived from Heisenberg equation yields undefined 
singularities, which depend on the initial conditions. The advantage of this 
new technique, based on Schwinger - Keldysh perturbation formalism, is that it 
explicitly determine these singular terms. Moreover, together with $EOM$ for 
retarded (advanced) Green's functions it allows to treat the problem in very 
consistent way making similar approximations in the decoupling procedure for 
all types of the Green's functions. Such approach was recently proposed to 
calculate the charge on the quantum dot upon nonequilibrium conditions 
\cite{MKKIWSSC}.

According to the Ref.\cite{Niu} the equation for the 'lesser' Green's functions
reads:
\begin{eqnarray}
\langle\langle A | B \rangle\rangle^<_{\omega} = 
g^<(\omega) \langle [A, B]_{\pm} \rangle + 
\nonumber \\
g^r(\omega) \langle\langle [A, H_I]_- | B \rangle\rangle^<_{\omega} +
g^<(\omega) \langle\langle [A, H_I]_- | B \rangle\rangle^a_{\omega}
\label{EOMNEQ}
\end{eqnarray}
where $g^{r (<)}(\omega)$ is the free electron $GF$ and $H_I$ denotes 
interacting part of the Hamiltonian. In general this equation allows to
calculate the the $GF$ on the left hand side, however in practice we have to 
approximate the higher order $GF$-s appearing on the right hand side (as in 
$EOM$ for equilibrium $GF$-s). But performing analogous approximations in the
decoupling procedure for both retarded and 'lesser' $GF$ we make theory
consistent. 

Applying above equation for the Green's function occurring in the formula for 
the current (\ref{CURR1}) $G^<_{N {\bf k} \sigma, d}(\omega)$, we get following
expression:
\begin{eqnarray}
\langle\langle \Psi_{N {\bf k} \sigma} | 
\Phi^+_{\sigma} \rangle\rangle^<_{\omega} =
g^r_{N {\bf k}}(\omega) H^I_{N {\bf k}}
\langle\langle \Phi_{\sigma} | \Phi^+_{\sigma} \rangle\rangle^<_{\omega} +
\nonumber\\
g^<_{N {\bf k}}(\omega) H^I_{N {\bf k}}
\langle\langle \Phi_{\sigma} | \Phi^+_{\sigma} \rangle\rangle^a_{\omega}
\label{EOMG<}
\end{eqnarray}
where as an interacting part of Hamiltonian $H_I$ we have taken the third term
in the Eq.(\ref{HAND}). $g^{r(<)}_{N {\bf k}}(\omega)$ is retarded ('lesser')
free-electron matrix $GF$ of the normal-state electrode.
\begin{eqnarray}
g^r_{N {\bf k}}(\omega) = 
\left( 
\begin{array}{cc}
\frac{1}{\omega - \epsilon_{N {\bf k}} + i0} & 0 \\
0 & \frac{1}{\omega + \epsilon_{N {\bf k}} + i0} 
\end{array}
\right)
\label{GF0RN}
\end{eqnarray}
\begin{eqnarray}
g^<_{N {\bf k}}(\omega) = 
\left( 
\begin{array}{cc}
2 \pi i f(\omega - eV) \delta(\omega - \epsilon_{N {\bf k}}) & 0 \\
0 & 2 \pi i f(\omega + eV) \delta(\omega + \epsilon_{N {\bf k}})
\end{array}
\right)
\label{GF0<N}
\end{eqnarray}
where $f(\omega)$ is the Fermi distribution function and $eV = \mu_N - \mu_S$
corresponds to the applied voltage between normal state electrode with the
chemical potential $\mu_N$ and superconducting one with $\mu_S$. In the
following we fix the chemical potential of the $SC$ electrode  ($\mu_S = 0$) 
and use $eV$ as a measure of the bias voltage. 

Expression (\ref{EOMG<}) is general for Anderson model and doesn't depend 
explicitly on the form of the Hamiltonian describing quantum dot ($H_{QD}$). 
The dependence enters only through the Green's function 
$G^{< (a)}_{d\sigma}(\omega) = 
\langle\langle \Phi_{\sigma} | \Phi^+_{\sigma} \rangle\rangle^{< (a)}_{\omega}$.
In the following we wish to study quantum dot in the limit of strong on-dot
Coulomb repulsion ($U_d \rightarrow \infty$). In this limit double occupancy of
the dot is forbidden and it is convenient to work in the slave boson 
representation in which the real electron operator is replaced by product of 
fermion and boson ones ($d_{\sigma} \Rightarrow b^+ f_{\sigma}$) 
\cite{Barnes,Coleman}. Additionally the fact that there is no double occupancy 
on the dot should be taken into account in some way. Usually such constraint 
is added to the Hamiltonian by the Lagrange multiplier. There is a number 
of variants of this approach in the literature and here we shall work within 
La Guillou - Ragoucy scheme \cite{LeGuillouRagoucy,MKKIWPRB}. In this approach 
the constraint of no double occupancy is enforced through modification of the 
commutation relations of both fermion and boson operators in comparison to 
the standard ones. This approach was successfully used in the study of the 
charge on the quantum dot \cite{MKKIWSSC}.

Hamiltonian of the system in the limit $U_d \rightarrow \infty$ in 
the slave boson representation is given in the form (\ref{HAND}), but now with 
\begin{eqnarray}
\Phi_{\sigma} = \left(
\begin{array}{l}
b^+ f_{\sigma} \\
f^+_{-\sigma} b 
\end{array}
\right)
\label{PHISB}
\end{eqnarray}
and 
\begin{eqnarray}
H^{QD} = \left(
\begin{array}{cc}
E_d & 0 \\
0 & - E_d
\end{array}
\right) 
\label{HQDSB}
\end{eqnarray}

Having introduced slave boson representation, we can begin calculations of the 
advanced and 'lesser' on-dot $GF$s appearing in Eq. (\ref{EOMG<}). To do this 
we apply formula (\ref{EOMNEQ}) together with usual prescription for the 
advanced (retarded) $GF$. One can investigate that on the higher-order $GF$s 
appeared in the this process have similar form in both cases: 'lesser' and 
advanced. So idea is to make the same approximations in the procedure of 
decoupling of the higher order $GF$s. Explicitly, we have performed decoupling
\begin{eqnarray}
\langle\langle c^+_{\lambda {\bf k} \sigma} c_{\lambda' {\bf k}' \sigma} A | 
B \rangle\rangle_{\omega} \approx 
\delta_{\lambda \lambda'} \delta_{{\bf k k}'}
n_{\lambda {\bf k}} \langle\langle A | B \rangle\rangle_{\omega}
\label{factornk}
\end{eqnarray}
and neglected the other $GF$s. In above formula $n_{\lambda {\bf k}}$ is the 
concentration of the electrons in the lead $\lambda$ in state ${\bf k}$ and for
superconducting electrode is given by
\begin{eqnarray}
n_{S {\bf k}} = 
\frac{1}{2} \left[ 1 - \frac{\epsilon_{S {\bf k}}}{E_{S {\bf k}}} 
\left(1 - 2 f(E_{S {\bf k}})\right)\right]
\label{nkS}
\end{eqnarray}
with quasiparticle spectrum
$E^2_{S {\bf k}} = \epsilon^2_{S {\bf k}} + \Delta^2_S$, while for 
the normal lead this relation reduces to 
\begin{eqnarray}
n_{N {\bf k}} =  f(\epsilon_{N {\bf k}}-eV)
\label{nkN}
\end{eqnarray}
We want to stress here, that we haven't used factorization like
\begin{eqnarray}
\langle\langle c^+_{\lambda {\bf k} \sigma} c^+_{\lambda' {\bf k}' \sigma} A |
B \rangle\rangle_{\omega} \approx 
\delta_{\lambda \lambda'} \delta_{{\bf k} - {\bf k}'} 
\langle c^+_{\lambda {\bf k} \sigma} c^+_{\lambda -{\bf k} \sigma} \rangle 
\langle\langle A | B \rangle\rangle_{\omega}
\label{factrodeltak}
\end{eqnarray}
The reason comes from the requirement of the hermicity relation between 
retarded and advanced Green's function, i. e.. 
$G^r(\omega) = [G^a(\omega)]^{\dagger}$. If we calculate retarded $GF$ within
$EOM$ and perform the same decoupling as in advanced Green's function keeping 
also 
$\langle c^+_{\lambda {\bf k} \sigma} c^+_{\lambda -{\bf k} \sigma} \rangle$
terms, we get expressions for the $GF$s which violates the hermicity relation. 
The only way to fulfill that at this level is to make approximations due to 
(\ref{factornk}) and neglect the remaining higher order $GF$s.

The resulting advanced on-dot $GF$  $G^a_{d\sigma}(\omega)$ can be written in 
the form of the Dyson equation:
\begin{eqnarray}
G^a_{d\sigma}(\omega) = g^a_{d\sigma}(\omega) +
g^a_{d\sigma}(\omega) \Sigma^a_{d\sigma}(\omega) G^a_{d\sigma}(\omega)
\label{DYSADV}
\end{eqnarray}
where $g^a_{d\sigma}(\omega)$ non-perturbed dot's advanced Green's function:
\begin{eqnarray}
g^a_{d\sigma}(\omega) = 
\left( 
\begin{array}{cc}
\frac{1 - n_{-\sigma}}{\omega - E_d - i0} & 0 \\
0 & \frac{1 - n_{\sigma}}{\omega + E_d - i0} 
\end{array}
\right)
\label{GF0AD}
\end{eqnarray}
and self energy $\Sigma^a_{d\sigma}(\omega)$ which can be written as sum of the
noninteracting $\Sigma^{0 a}_d(\omega)$ and interacting 
$\Sigma^{I a}_d(\omega)$ part 
\begin{eqnarray}
\Sigma^a_d(\omega) = \Sigma^{0 a}_d(\omega) + \Sigma^{I a}_d(\omega) = 
\sum_{\lambda {\bf k}} 
\left[\Sigma^{0 a}_{\lambda {\bf k}}(\omega) + 
\Sigma^{I a}_{\lambda {\bf k}}(\omega)\right]
\label{SIGTOT}
\end{eqnarray}
For superconducting electrode we have
\begin{eqnarray}
\fl
\Sigma^{0 a}_{S {\bf k}}(\omega) =
\left(
\begin{array}{cc}
\frac{|V_{S {\bf k}}|^2}{(1 - n_{-\sigma})}
\left(\frac{u^2_{S {\bf k}}}{\omega - E_{S {\bf k}} - i0} +
\frac{v^2_{S {\bf k}}}{\omega + E_{S {\bf k}} - i0}\right) 
&
\frac{|V_{S {\bf k}}|^2}{(1 - n_{\sigma})}
\left(\frac{u_{S {\bf k}} v_{S {\bf k}}}
{\omega + E_{S {\bf k}} - i0} -
\frac{u_{S {\bf k}} v_{S {\bf k}}}
{\omega - E_{S {\bf k}} - i0}\right) \\
\\
\frac{|V_{S {\bf k}}|^2}{(1 - n_{-\sigma})}
\left(\frac{u_{S {\bf k}} v_{S {\bf k}}}
{\omega + E_{S {\bf k}} - i0} -
\frac{u_{S {\bf k}} v_{S {\bf k}}}
{\omega - E_{S {\bf k}} - i0}\right)
&
\frac{|V_{S {\bf k}}|^2}{(1 - n_{\sigma})}
\left(\frac{v^2_{S {\bf k}}}{\omega - E_{S {\bf k}} - i0} +
\frac{u^2_{S {\bf k}}}{\omega + E_{S {\bf k}} - i0}\right)\\
\end{array}
\right)
\label{SIGMA0S}
\end{eqnarray}
where we have introduced $BCS$ factors 
$u^2_{S {\bf k}} = 
\frac{1}{2} \left(1 + \frac{\epsilon_{S {\bf k}}}{E_{S {\bf k}}}\right)$, 
$v^2_{S {\bf k}} = 
\frac{1}{2} \left(1 - \frac{\epsilon_{S {\bf k}}}{E_{S {\bf k}}}\right)$. For 
the normal state corresponding expression is:
\begin{eqnarray}
\Sigma^{0 a}_{N {\bf k}}(\omega) =
\left(
\begin{array}{cc}
\frac{|V_{N {\bf k}}|^2}{(1 - n_{-\sigma})}
\frac{1}{\omega - \epsilon_{N {\bf k}} - i0}
& 0 \\
0 &
\frac{|V_{N {\bf k}}|^2}{(1 - n_{\sigma})}
\frac{1}{\omega + \epsilon_{N {\bf k}} - i0}\\
\end{array}
\right)
\label{SIGMA0N}
\end{eqnarray}
It turns out that within the present approach the interacting part of the self 
energy is simply related to the noninteracting one. Moreover the same relation
also holds for retarded as well as 'lesser' $GF$s. This is a result of the
consistency of the decoupling procedure and requirement of the hermicity 
relation between retarded and advanced $GF$. In general this relation can be 
written as:
\begin{eqnarray}
\Sigma^{I}_{\lambda {\bf k}}(\omega) = 
n_{\lambda {\bf k}} \tau_3 \Sigma^{0}_{\lambda {\bf k}}(\omega) \tau_3
\label{SIGMAINT}
\end{eqnarray}
where $\tau_3 = \left(
\begin{array}{cr}
1 & 0 \\
0 & -1 \\
\end{array}\right)$ 
is the Pauli matrix and $n_{\lambda {\bf k}}$ is the concentration of
the electrons of the wave vector ${\bf k}$ in the lead $\lambda$ given by 
(\ref{nkS}) and (\ref{nkN}).

It is possible to write Eq. for the 'lesser' $GF$ in the form of the Keldysh
equation (\ref{KELDEQ}) with $G^a_{d\sigma}(\omega)$ given by Eq.(\ref{DYSADV})
and $G^r = [G^a]^{\dagger}$. Free electron dot's 'lesser' Green's function is 
given in the form:
\begin{eqnarray}
g^<_{d\sigma}(\omega) = 
\left( 
\begin{array}{cc}
2 \pi i (1 - n_{-\sigma}) f(\omega) \delta^-_d & 0 \\
0 & 2 \pi i (1 - n_{\sigma}) f(\omega) \delta^+_d
\end{array}
\right)
\label{GF0<D}
\end{eqnarray}
where $\delta^{\pm}_d = \delta(\omega \pm E_d)$. As we have mentioned, the 
'lesser' self energy has the same form as advanced 
one: 
\begin{eqnarray}
\Sigma^<_d(\omega) = \Sigma^{0 <}_d(\omega) + \Sigma^{I <}_d(\omega) = 
\sum_{\lambda} 
(\Sigma^{0 <}_{\lambda{\bf k}}(\omega) + 
\Sigma^{I <}_{\lambda {\bf k}}(\omega))
\label{SIGTOT<}
\end{eqnarray}
where noninteracting part due to $SC$ lead is:
\begin{eqnarray}
\fl
\Sigma^{0 <}_{S {\bf k}}(\omega) =
2 \pi i f(\omega) 
\left(
\begin{array}{cc}
\frac{|V_{S {\bf k}}|^2}{(1 - n_{-\sigma})} 
\left(u^2_{S {\bf k}}\delta^-_S +
v^2_{S {\bf k}}\delta^+_S\right) 
&
\frac{|V_{S {\bf k}}|^2}{(1 - n_{\sigma})} 
u_{S {\bf k}} v_{S {\bf k}} \left(
\delta^+_S - \delta^-_S\right) \\
\\
\frac{|V_{S {\bf k}}|^2}{(1 - n_{-\sigma})}
u_{S {\bf k}} v_{S {\bf k}}\left(
\delta^+_S - \delta^-_S\right)
&
\frac{|V_{S {\bf k}}|^2}{(1 - n_{\sigma})}
\left(v^2_{S {\bf k}}\delta^-_S +
u^2_{S {\bf k}}\delta^+_S\right)\\
\end{array}
\right) 
\label{SIGMA0S<}
\end{eqnarray}
and for the normal lead we have
\begin{eqnarray}
\Sigma^{0 <}_{N {\bf k}}(\omega) =
2 \pi i \left(
\begin{array}{cc}
\frac{|V_{N {\bf k}}|^2}{(1 - n_{-\sigma})}
f(\omega - eV) \delta^-_N
& 0 \\
0 &
\frac{|V_{N {\bf k}}|^2}{(1 - n_{\sigma})}
f(\omega + eV) \delta^+_N\\
\end{array}
\right)
\label{SIGMA0N<}
\end{eqnarray}
where $\delta^{\pm}_S = \delta(\omega \pm E_{S{\bf k}})$ and 
$\delta^{\pm}_N = \delta(\omega \pm \epsilon_{N{\bf k}})$. 
And again, the interacting part of the 'lesser' self energy 
is related to the noninteracting one simply through Eq.(\ref{SIGMAINT}). 

Now we are ready to write the expression for the current (\ref{CURR1}) in terms
of known $GF$s. First, let's rewrite the Keldysh equation for the element $11$
of the dot's $GF$ in the form:
\begin{eqnarray}
G^<_{11} = G^r_{11} \Sigma^<_{11} G^a_{11} + G^r_{11} \Sigma^<_{12} G^a_{21} +
\nonumber \\
G^r_{12} \Sigma^<_{21} G^a_{11} + G^r_{12} \Sigma^<_{22} G^a_{21}
\label{GD<11}
\end{eqnarray}
Note that we don't have term proportional to $g^<_d$ as it vanishes in our 
case \cite{Yauho}.

To calculate the $GF$ given by Eq.(\ref{EOMG<}), entering to the expression for
the current (\ref{CURR1}), we need yet element $11$ of the advanced $GF$, more
precisely imaginary part of that. Note that $g^<_{N {bf k}}$ is purely imaginary
and we need real part of $g^<_{N {\bf k}} G^a_d$. We can write down equation 
for the imaginary part of the element $11$ of $G^a_d$ in the similar fashion as 
Eq.(\ref{GD<11}), i.e.
\begin{eqnarray}
{\it Im} G^a_{11} = G^r_{11} {\it Im} \Sigma^a_{11} G^a_{11} + 
G^r_{11} {\it Im} \Sigma^a_{12} G^a_{21} +
\nonumber \\
G^r_{12} {\it Im} \Sigma^a_{21} G^a_{11} + 
G^r_{12} {\it Im} \Sigma^a_{22} G^a_{21}
\label{GDA11}
\end{eqnarray}

Substituting now the Eqs. (\ref{GD<11}) and (\ref{GDA11}) into (\ref{EOMG<}) we
get expression for the $G_{N {\bf k}, d}$, which determines the current
(\ref{CURR1}). Finally the current (\ref{CURR1}) can be written as
\begin{eqnarray}
J = J_{11} + J_{22} + J_{12} + J_A
\label{JTOT}
\end{eqnarray}
The first term represents conventional tunneling and is given in the form
\begin{eqnarray}
J_{11} = - \frac{2e}{\hbar} 
\int^{\infty}_{-\infty} \frac{d\omega}{2\pi} {\it Im} \Sigma^S_{11} |G_{11}|^2
\nonumber \\
\Gamma_N \rho^N_{11} [f(\omega - eV) - f(\omega)]
\label{J11}
\end{eqnarray}
where is the elastic rate defined as $\Gamma_N = 2 \pi V^2_{N} \rho^N(0)$ and 
$\rho^N(0)$ is the bare normal state density of states at the Fermi energy.
The second term describes the 'branch crossing' process (process with crossing
through the Fermi surface) in the language of the $BTK$ theory (Blonder -
Tinkham - Klapwijk) \cite{Blonder}: electron from the normal lead is converted
into the hole like in the $SC$ lead.
\begin{eqnarray}
J_{22} = - \frac{2e}{\hbar} 
\int^{\infty}_{-\infty} \frac{d\omega}{2\pi} {\it Im} \Sigma^S_{22} |G_{12}|^2
\nonumber \\
\Gamma_N \rho^N_{11} [f(\omega - eV) - f(\omega)]
\label{J22}
\end{eqnarray}
The next term corresponds to the process in which electron tunnels into $SC$
picking up the quasiparticle and creating a Cooper pair.
\begin{eqnarray}
J_{12} = - \frac{4e}{\hbar} 
\int^{\infty}_{-\infty} \frac{d\omega}{2\pi} 
{\it Im} \Sigma^S_{12} {\it Re}[G_{11} G^*_{12}]
\nonumber \\
\Gamma_N \rho^N_{11} [f(\omega - eV) - f(\omega)]
\label{J12}
\end{eqnarray}
The last term in (\ref{JTOT}) represents Andreev tunneling in which electron
from the normal lead is reflected back as a hole and Cooper pair is created in
superconducting electrode.
\begin{eqnarray}
J_{A} = - \frac{2e}{\hbar} 
\int^{\infty}_{-\infty} \frac{d\omega}{2\pi} {\it Im} \Sigma^N_{22} |G_{12}|^2
\nonumber \\
\Gamma_N \rho^N_{11} [f(\omega - eV) - f(\omega + eV)]
\label{JA}
\end{eqnarray}

As one can see, at energies $|eV| < \Delta_S$ and zero temperature, the only
process which contribute to the total current, is the Andreev tunneling. The
remaining ones represent 'single particle' processes which are suppressed at 
$|eV| < \Delta_S$ due to the lack of the states in superconductor. Of course for
energies $|eV| > \Delta_S$ all these processes give rise to the current, even
Andreev does, however is strongly suppressed, but still finite. It is also
worthwhile to note that all these processes (except $J_{11}$) proceed through
virtual states on the dot.


\section{\label{sec3} Density of states}

In the following sections we will present numerical results of electron
tunneling in the $N - QD -S$ system and show how different terms of Eq.
(\ref{JTOT}) contribute to the total current and differential conductance. But
firstly we want to turn our attention to the density of states as it gives a 
lot of information about system.

The most pronounced fingerprint of the Kondo effect in the $N - QD N$ system is
the Abrikosov - Suhl or Kondo resonance at the Fermi level and its temperature
dependence. Kondo resonance appears as temperature is lower than parameter 
dependent Kondo temperature $T_K$. In the original Kondo effect there is odd
number of electrons on the dot, so the total spin is half - integer. In this 
case electrons from the leads with energy close to Fermi level screen the spin 
on the dot producing resonance at the Fermi energy. If electrodes are made
superconducting situation is more complicated, as there enters another energy
scale - superconducting transition temperature $T_c$ (or equivalently $SC$
order parameter $\Delta$). And the Kondo effect takes place provided 
$T_K > T_c$, otherwise is absent due to lack of the low energy states in the
leads to screen the spin on the dot. Naturally there raises a question what 
will happen if one of the lead is superconducting and another in the normal
state. This was investigated in \cite{Fazio}-\cite{Raimondi}, \cite{Clerk} and
it has been found that Kondo effect survives in the presence of
superconductivity in one of the electrodes, even $T_K < T_c$. The reason for 
this is simply: the spin of the dot is screened by electrons in the normal 
lead. We will show that this is a really the case and this is seen in the
density of states of quantum dot.

\begin{figure}[h]
\begin{center}
 \resizebox{8.4cm}{!}{
  \includegraphics{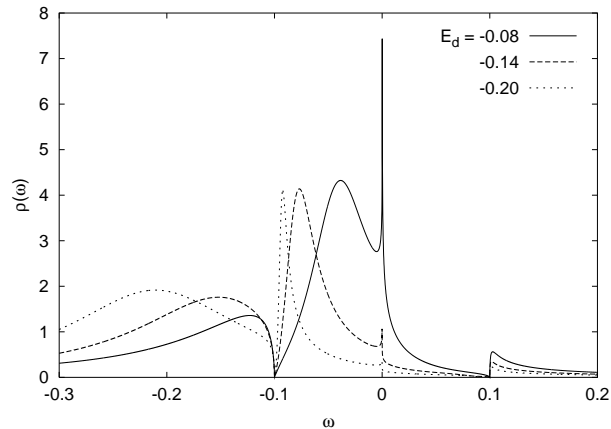}}
\end{center}
 \caption{\label{Fig1} The density of states of the quantum dot for various 
          values of the dot energy level $E_d$. Other parameters are 
	  following: $\Gamma_N  = \Gamma_S = 0.02$, $\Delta = 0.1$, $eV = 0$, 
	  $T = 10^{-5}$ in the units of the bandwidth $W$.}
\end{figure}

In the Fig.\ref{Fig1} we show the density of states of the quantum dot for
various positions of the dot energy level $E_d$. 
It is clearly seen that Kondo
effect, which manifests itself in the resonance on the Fermi level, survives 
the presence of superconductivity in one electrode. The additional structure at
$\omega = \pm \Delta$ coming from the $SC$ lead is also visible. This is 
simply reflection of the $SC$ gap. At this point it worthwhile to note, that if
$|E_d| > \Delta$ there is bound-like (Andreev) state within the $SC$ gap,
position of which depends on $E_d$. In the $S - QD - S$ system this is a true
bound state. However in the present case, due to finite $DOS$ in the normal 
lead, this state acquires a finite width (resonance state). 

\begin{figure}[h]
\begin{center}
 \resizebox{8.4cm}{!}{
  \includegraphics{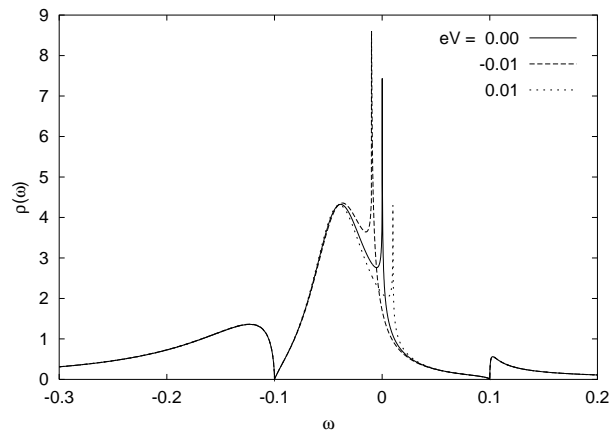}}
\end{center}
\caption{Equilibrium ($eV=0$) and nonequilibrium ($eV = \pm 0.01$) density of 
states of the quantum dot. Other parameters have following values: 
$\Gamma_N = \Gamma_S = 0.02$, $\Delta = 0.1$, $E_d = -0.08$, $T = 10^{-5}$ in 
units of the bandwidth $W$.}
\label{Fig2}
\end{figure}

It is very interesting to see how the $DOS$ will be look like in the
nonequilibrium situation ($eV = \mu_N - \mu_S = \neq 0$). Let us recall that 
the Kondo resonance is located at the Fermi level of the lead. In the 
$N - QD - N$ system when $eV \neq 0$ there emerge two resonances at Fermi 
levels of the left and right lead respectively. In our case there is a gap in
the $SC$ lead $DOS$, and if our simple picture that Kondo effect is only due to
normal lead, we expect only one resonance pinned to the normal metal electrode
Fermi level ($\mu_N$). As we can learn from the Fig.\ref{Fig2} this is really
the case - the Kondo resonance follows the Fermi level of the normal lead.

\begin{figure}[h]
\begin{center}
 \resizebox{8.4cm}{!}{
  \includegraphics{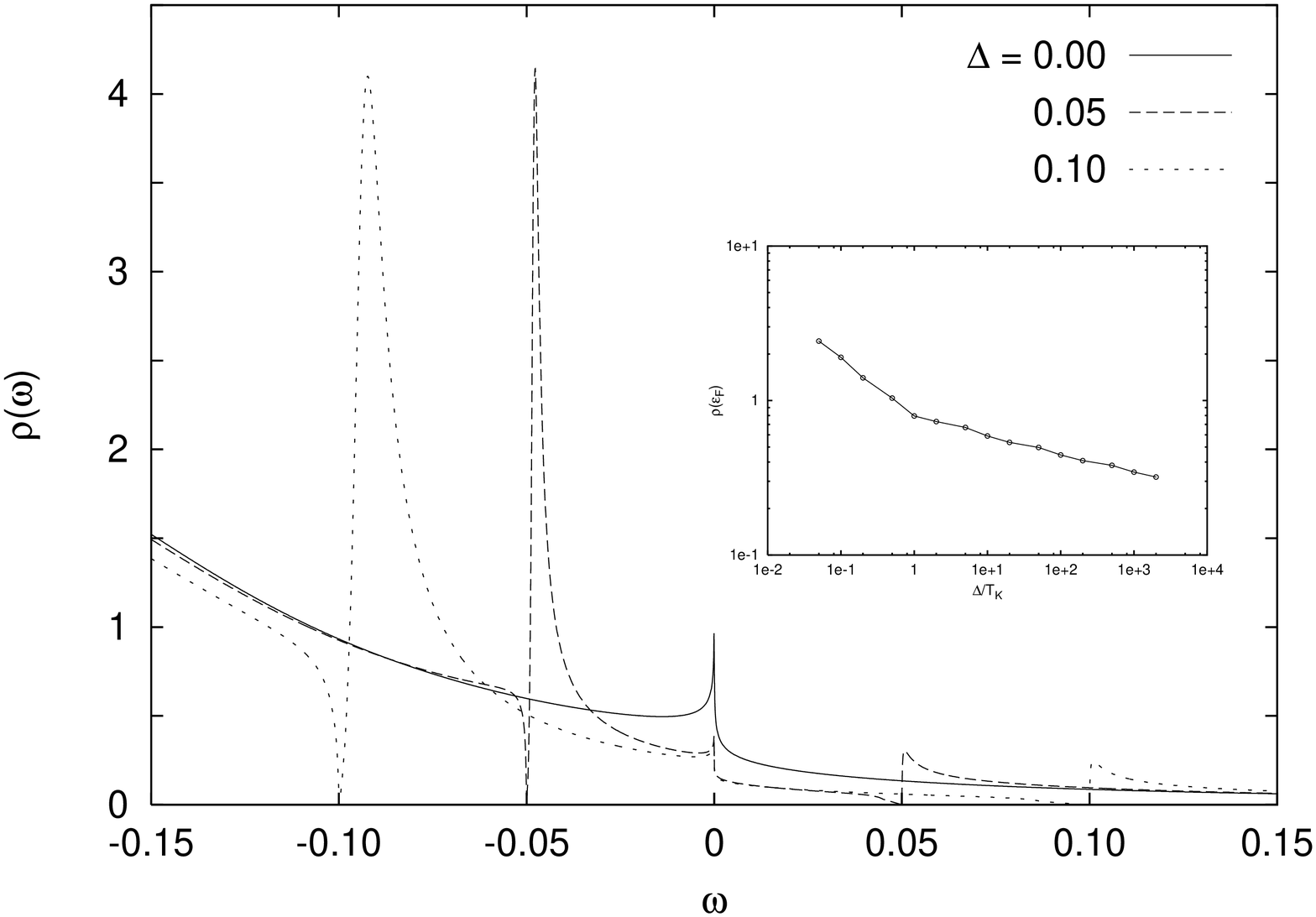}}
\end{center}
\caption{Equilibrium density of states of the quantum dot for various values of 
the order parameter ($\Delta$). $E_d=-0.2$ and other parameters have values as 
in the Fig.(\ref{Fig2}).}
\label{Fig3}
\end{figure}

In the Fig.\ref{Fig3} the $DOS$ is plotted for a few values of the $SC$ order
parameter $\Delta$. 
As we can see the Kondo resonance is strongly suppressed in comparison to the 
$N - QD - N$ $DOS$ (solid line). The reason for that is that due to lack of the 
low laying states in $SC$, the spin on the dot is weakly screened. Similar 
conclusions have been reached by A. A. Clerk and coworkers \cite{Clerk} within 
$NCA$ approach. In the inset the density of states at the Fermi level 
$\rho(\varepsilon_F)$ is plotted as a function of the $SC$ order parameter 
$\Delta$. 

\begin{figure}[h]
\begin{center}
 \resizebox{8.4cm}{!}{
  \includegraphics{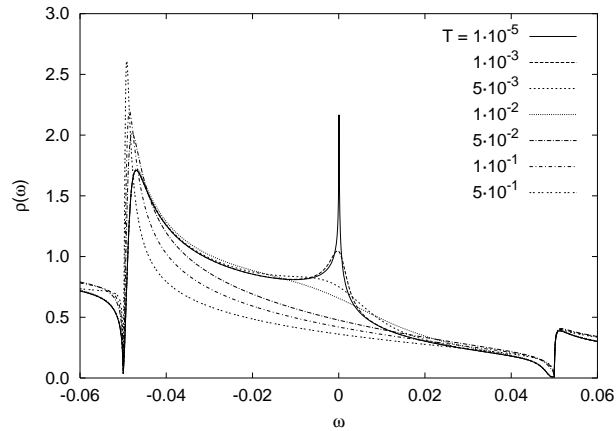}}
\end{center}
\caption{Temperature dependence of the $DOS$ for $\Gamma_N = 0.05$, 
$\Gamma_S = 0.02$, $\Delta = 0.05$, $E_d = -0.2$, $eV = 0$.}
\label{Fig4}
\end{figure}

\begin{figure}[h]
\begin{center}
 \resizebox{8.4cm}{!}{
  \includegraphics{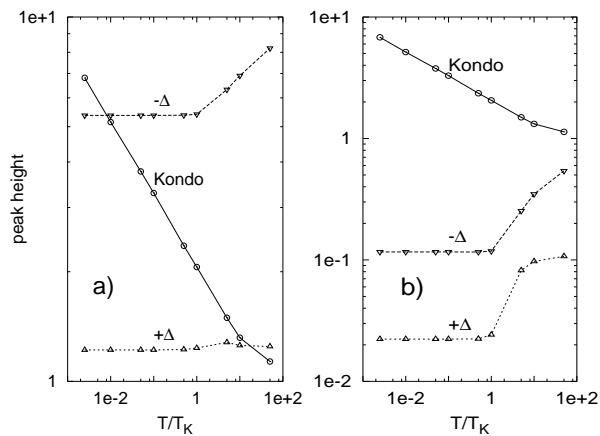}}
\end{center}
\caption{Temperature dependence of height of the peaks $(a)$ and dips $(b$) 
near $\pm \Delta$ (see
Fig.(\ref{Fig3})). For comparison the temperature dependence of the Kondo 
resonance is also shown. The parameters are the same as in the 
Fig(\ref{Fig3}).}
\label{Fig5}
\end{figure}

Finally we want to discuss the temperature dependence of the dot density of
states. In the Fig.\ref{Fig4} we show the $DOS$ for number of temperatures. 
As we expected the Kondo resonance disappears as temperature is raised. 
However it is important to stress out that resonances and dips 
$\omega \approx \Delta$ are also temperature dependent. It is clearly seen in 
the Fig.\ref{Fig5}, where height of the peaks $(a)$ and dips $(b)$ are shown. 
The spectral weight of the
Kondo resonance is also shown for comparison. It is worthwhile to note that
below $T_K$ the height of both peaks and dips are constant. As soon as
temperature exceeds $T_K$ height of these resonances starts to raise as well as
dips does. This effect can be though of as a transfer of the spectral weight
between Kondo resonance and Andreev states. Let us remind that Andreev
processes still take place at $T > T_K$ but less than $T_c$. We want to notice
that our results are in contradiction to what has been found in
Ref.\cite{Clerk} within $NCA$. The authors of \cite{Clerk} have shown that the
resonances at $\pm \Delta$ do not appear at higher temperatures. This has a
important consequences. This means that, even for $T_c > T_K$, superconducting
electrons do participate in the Kondo effect. So our simple picture breaks 
down.

As we mentioned, they applied $NCA$ technique to calculate the dot $DOS$, which
is known to give correct results in the normal state for a wide range of
temperatures. It is also known, that $EOM$ gives quantitatively incorrect
results at low temperatures ($T < T_K$). But here this effect certainly take
place at $T > T_K$ in the range of validity of $EOM$. So the question of which
picture is indeed realized in the real system seems to remain open.


\section{\label{sec4} Andreev reflections and the Kondo effect}

We have shown, that Kondo peak in the density of states survives the presence 
of the superconductivity, however should we expect that peak in the 
current - voltage characteristic (differential conductance - 
$G(eV) = dJ/d(eV)$) ? If we consider tunneling processes, described 
by Eqs. (\ref{JTOT})-(\ref{JA}), we might expect Kondo peak only in the 
tunneling mediated by the Andreev reflections (\ref{JA}). The amplitudes of 
the other processes is equal to zero (at $T=0$) for energies less than $SC$ 
gap. 

\begin{figure}[h]
\begin{center}
 \resizebox{8.4cm}{!}{
  \includegraphics{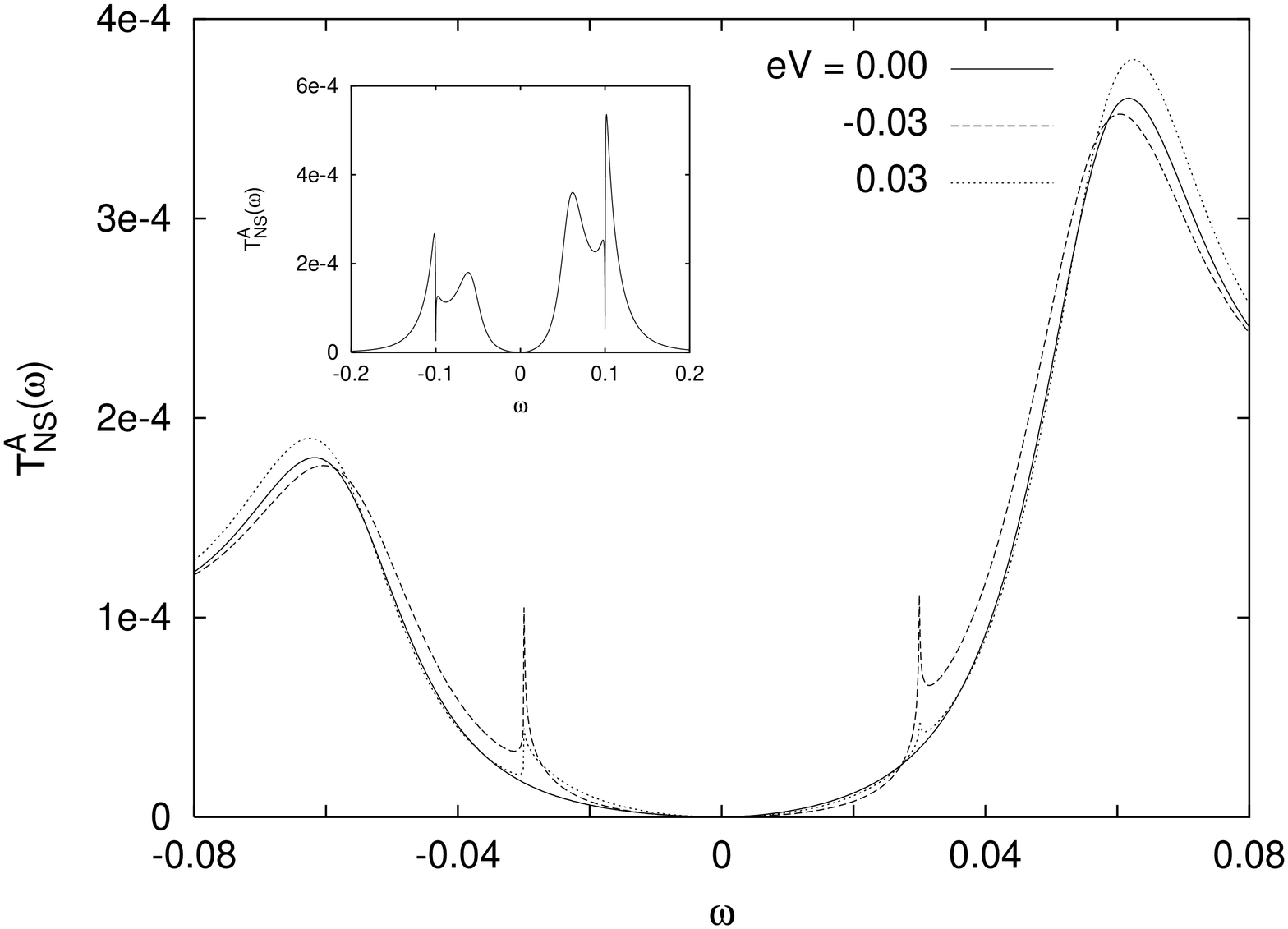}}
\end{center}
\caption{$T^A_{NS}(\omega)$ for different values of the bias voltage $eV=0$ 
(solid line), $-0.03$ (dashed) and $0.03$ (dotted line). $E_d = -0.08$, 
$\Gamma_N = \Gamma_S = 0.01$ and $\Delta = 0.1$. Inset: large scale view of the 
equilibrium $T^A_{NS}(\omega)$.}
\label{Fig6}
\end{figure}

Let's rewrite Eq. (\ref{JA}) into the form:
\begin{eqnarray}
J_A = - \frac{2 e}{h} \int^{\infty}_{-\infty} \frac{d\omega}{2\pi}
T^A_{NS}(\omega) [f(\omega - eV) - f(\omega + eV)]
\label{JAGNS}
\end{eqnarray}
We have introduced 'transmittance' $T^A_{NS}(\omega)$, associated with the
Andreev tunneling, defined as:
\begin{eqnarray}
T^A_{NS}(\omega) = \frac{e}{\hbar} \Gamma^2_N \rho^N_{11}(\omega) 
\rho^N_{22}(\omega) |G_{21}(\omega)|^2
\label{GNS}
\end{eqnarray}
In fact, at zero temperature and at energies less than superconducting gap 
$T^A_{NS}(\omega)$ can be regarded as a total transmittance, because Andreev 
tunneling is only process allowed in these circumstances. $T^A_{NS}(\omega)$ 
for different values of the $eV$ is plotted in the Fig. \ref{Fig6}.
The broad resonances at $\omega \approx \pm 0.06$ are reflections of the dot 
energy level $E_d = 0.08$ for electrons and holes \cite{Fazio}, shifted from 
its original position due to renormalization caused by the strong Coulomb
interaction. But more important point is that there is no Kondo peak in 
equilibrium ($eV = 0$) transmittance. This is in agreement with Refs.
\cite{Fazio,Clerk}. This is because the imaginary part of the anomalous 
Green's function $G_{12}(\omega)$ behaves like $|\omega|$ for 
$\omega \ll \Delta$ while its real part is proportional to 
$\omega ln{(\omega)}$ and both vanish for $\omega = 0$. And this is sufficient
to suppress the Kondo effect. However as soon as we go away from the $eV = 0$,
we can observe the Kondo peaks at energies $\omega = \pm eV$ with 
approximately equal spectral weight. However there is strong asymmetry between
negative (dashed line) and positive (dotted line) voltages. While in former 
case we have very well resolved resonances, in the later these resonances are
strongly suppressed. This asymmetry is strictly related to the density of 
states (see Fig. (\ref{Fig2})), where we also observe such asymmetry, which is
associated with different conditions for the Kondo effect in both cases (note
quantity $E_d - eV$). The fact, that we observe the Kondo peak for both 
electrons ($eV$) and for holes ($- eV$) is in contradiction to what has been 
observed in Ref. \cite{Fazio}, where only small kink has emerged for 
$\omega = - eV$. This is certainly due to different approximation scheme used 
in calculations.

\begin{figure}[h]
\begin{center}
 \resizebox{8.4cm}{!}{
  \includegraphics{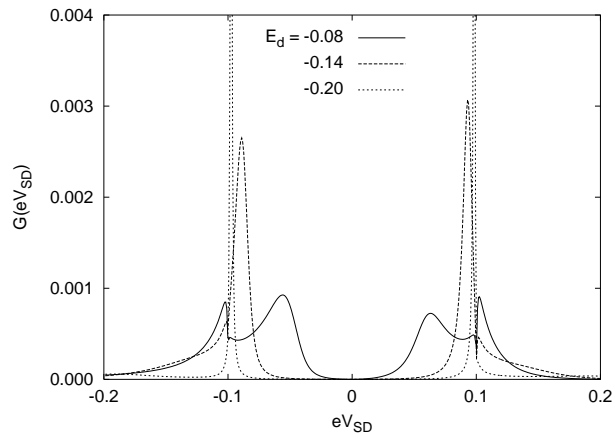}}
\end{center}
\caption{The Andreev differential conductance 
         $G_A(eV_{SD}) = d J_A/d (eV_{SD})$ for different values of the dot
	 energy level $E_d=-0.08$ (solid line), $-0.14$ (dashed) and $0.2$ 
	 (dotted line). $\Gamma_N = \Gamma_S = 0.01$ and $\Delta = 0.1$.}
\label{Fig7}
\end{figure}

Since equilibrium transmittance $T^A_{NS}(\omega)$ doesn't show the Kondo peak,
we cannot expect it in the differential conductance 
$G_A(eV_{SD}) = d J_A/d (eV_{SD})$ with $J_A$ defined by (\ref{JA}), since 
$G_A$ is proportional to the equilibrium $T^A_{NS}$. Indeed this is what we
observe in the (Fig. \ref{Fig7}). 
We see, that $G_A(eV_{SD})$ is very sensitive to the position of the dot 
energy level. The larger (negative) $E_d$ the smaller conductance. It can be 
understood as follows. The probability of the Andreev reflections depends on 
the density of states (for $\omega < \Delta$) of the normal electrode as well 
as dot itself. The later one is strongly $E_d$-dependent (see Fig. \ref{Fig1}). 
For $E_d \ll 0$ there are no states on the dot participated in tunneling 
between normal electrode and the superconductor. Note that in fact Andreev 
reflections take place between $SC$ electrode and the dot.

The lack of the peak in the differential conductance confirms that the Kondo 
effect is suppressed in the $N - QD - S$ system with strong on-dot Coulomb
repulsion. This result is in full agreement with those of Refs. 
\cite{Fazio,Clerk,Cuevas}.


\section{\label{sec5} Asymmetric coupling}

The quantum dot asymmetrically coupled to the normal state electrodes shows
anomalous Kondo effect \cite{MKKIWPRB_2}, which has also been observed
experimentally \cite{Schmid,Simmel}. This anomaly features in the non-zero
position of the Kondo resonance in the differential conductance. In other words,
if we increase one of the couplings to the leads ($\Gamma_L(R)$), the zero-bias
anomaly moves to non-zero voltages \cite{MKKIWPRB_2}. 
In the $N - QD - S$ system, there is no Kondo resonance in the differential 
conductance. Similarly there is no it in the equilibrium transmittance. On the 
other hand the Kondo peak emerges when system is in nonequilibrium. So in fact 
we could expect the non-zero bias Kondo peak in the differential conductance of 
the dot asymmetrically coupled to the leads.

\begin{figure}[h]
\begin{center}
 \resizebox{8.4cm}{!}{
  \includegraphics{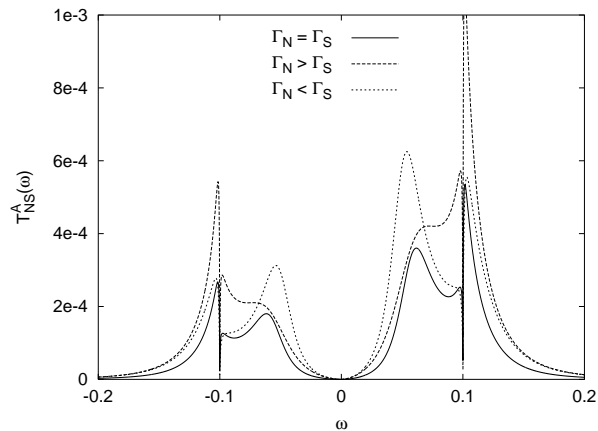}}
\end{center}
\caption{The Andreev transmittance $T^A_{NS}$ for different values of the
         couplings to the leads $\Gamma_N = \Gamma_S = 0.01$ (solid line), 
	 $\Gamma_N = 0.015$, $\Gamma_S = 0.01$ (dashed) and $\Gamma_N = 0.01$, 
	 $\Gamma_S = 0.015$ (dotted line). $E_d = - 0.08$ and $\Delta = 0.1$.}
\label{Fig8}
\end{figure}

We have calculated Andreev transmittance and differential conductance for a
number of couplings to the leads. The example is shown in the Fig. \ref{Fig8}.
Unfortunately we haven't observed the Kondo peak at non-zero voltages 
regardless how big the asymmetry ($\Gamma_N/\Gamma_S$) was. The reason for this
might be that for $N - QD - N$ system the shift of the Kondo peak to non-zero 
values is very small \cite{MKKIWPRB_2}, and in the present case the Kondo
resonance cannot develop because there are too few states in the transmittance
spectra around the Fermi energy.

However the asymmetry in the couplings lead to another interesting behavior of
the Andreev transmittance as well as differential conductance. Namely it turns
out that $\Gamma_N$ more influences the $T^A_{NS}$ ($G_A$) around energies 
close to value of $SC$ gap than $\Gamma_S$ does (see dashed line in the 
Fig. \ref{Fig8}). On the other hand the transmittance (conductance) around 
$|E_d| < \Delta$ is more affected by $\Gamma_S$ (compare dotted line in the 
Fig. \ref{Fig8}). We can also note that the positions of the broad resonances, 
corresponding to the dot energy level $E_d$, depend on asymmetry. This is due
to the real part of self-energy. More important is that $\Gamma_N$ and 
$\Gamma_S$ shift the positions of the resonances in opposite directions:
$\Gamma_N$ towards Fermi energy $\mu_N = \mu_S = 0$ while $\Gamma_S$ to higher 
energies.

\begin{figure}[h]
\begin{center}
 \resizebox{8.4cm}{!}{
  \includegraphics{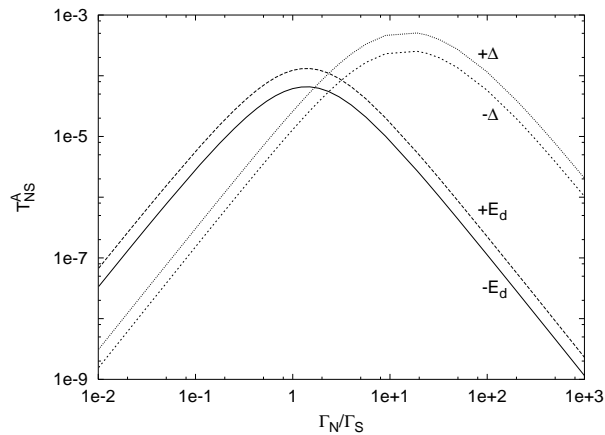}}
\end{center}
\caption{$T^A_{NS}$ at energies: $-E_d$ (solid line), $E_d$ (dashed),
         $-\Delta$ (dotted) and $\Delta$ (dot-dashed line) as function of the
	 $\Gamma_N/\Gamma_S$ for $E_d = -0.08$, $\Delta = 0.1$ and fixed
	 $\Gamma_N + \Gamma_S = 0.02$.}
\label{Fig9}
\end{figure}

In the present approach the Andreev transmittance (and differential 
conductance) vanishes at zero energy. However it shows interesting properties 
at the other characteristic energies of the system, like $E_d$ or $\Delta$. 
The Andreev transmittance $T^A_{NS}$ at these energies is shown in the 
Fig. \ref{Fig9} as a function of the asymmetry in the coupling
$\Gamma_N/\Gamma_S$.
We see that the tunneling due to the Andreev processes at energies $\pm E_d$ is
likely to take a place when the couplings are more or less symmetric, i. e. 
$\Gamma_N = \Gamma_S$. More surprising result is that the largest probability 
of these processes at energies $\pm \Delta$ is for large asymmetry 
$\Gamma_N/\Gamma_S \approx 10$. As we already mentioned such behavior can be
explained by renormalization of the dot energy level due to the real part of
self-energy, which depends on $\Gamma_N$ and $\Gamma_S$ in rather a complicated 
way .


\section{\label{sec6} Particle-hole asymmetry}

Until now we have presented results for the special case, namely the
electron-hole ($e-h$) symmetry in the leads. It is well known \cite{Hirsch} 
that the particle-hole asymmetry in the normal metal/superconductor tunnel 
junctions and metallic contacts suppresses the Andreev reflections due to the 
fact that the reflection and transmission probabilities are different for 
incident electrons and holes. In the quantum dot coupled to the normal and 
superconducting electrode we could also expect that this asymmetry will play 
a role. On the other hand, this asymmetry is already present in the strongly 
interacting quantum dot ($U = \infty$) in the Kondo regime ($E_d < 0$), as one 
studied here. However this asymmetry in the leads can further modify the 
Andreev tunneling.

\begin{figure}[h]
\begin{center}
 \resizebox{8.4cm}{!}{
  \includegraphics{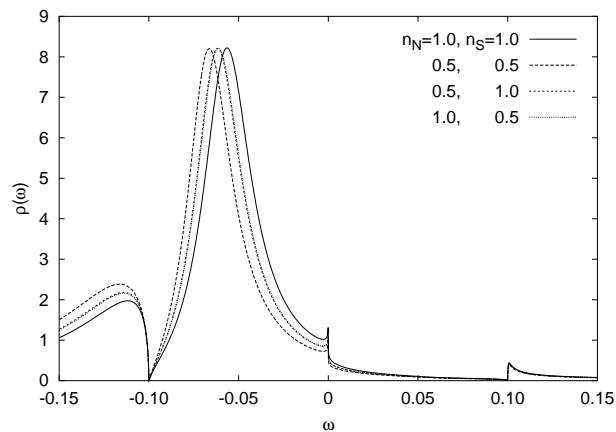}}
\end{center}
\caption{Density of states in various realizations of the electron-hole
         asymmetry indicated in the figure. Parameters have following values: 
	 $E_d = -0.08$, $\Delta = 0.01$, $\Gamma_N = \Gamma_S = 0.01$.}
\label{Fig10}
\end{figure}

Let's start from the effect of the $e-h$ asymmetry on the density of states. As
we can read from the Fig. \ref{Fig10}, the asymmetry plays rather a minor role.
The most pronounced difference is when the concentration in both $N$ and $S$
electrodes are changed. If change the concentration in one of the electrode, 
the effect is smaller. Moreover, the density of states almost does not depend 
on in which electrode concentration is changed. In other words, it seems to be
sensitive to average concentration in both electrodes. 

\begin{figure}[h]
\begin{center}
 \resizebox{8.4cm}{!}{
  \includegraphics{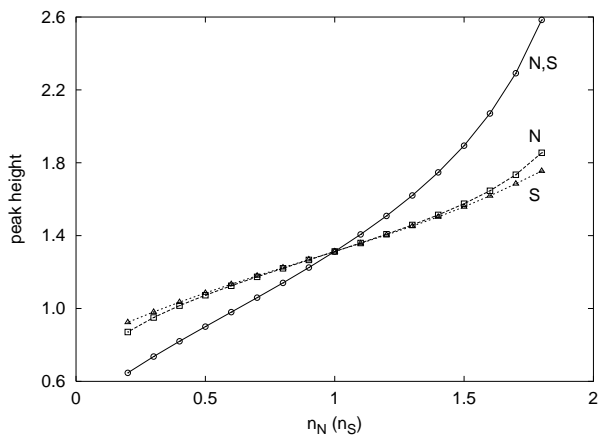}}
\end{center}
\caption{The height of the Kondo peak as a function of the concentrations of 
         the electrons $n_N$ ($n_S$) in the normal (superconducting) lead.
         Curve $N,S$ - electron concentration in both electrodes is changed, 
	 $N$ - in the normal electrode only and $S$ - in superconducting. The 
	 parameters are the same as in Fig. \ref{Fig10}.}
\label{Fig11}
\end{figure}

In the Fig. \ref{Fig11} we have shown the height of the Kondo resonance when
electron concentration is varied. 
As one can read from the Fig. \ref{Fig11}, the spectral weight of the Kondo 
peak strongly depends on the electron concentration in both electrodes. 
Moreover there is a strong asymmetry with respect to the $n = 1$ point, i. e. 
the peak is higher when the concentration of electrons is higher. It is rather
expected result, as in the original Kondo effect the resonance at zero energy
emerges due to the screening of the conduction electrons. So one could expect
that it should depend on their concentration, as it does. Similar effect we
observe when the electron concentration is changed in one lead only. It almost
does not depend on in which lead $n$ is changed. However for large $e-h$
asymmetry, concentration of electrons in the normal lead seems to play a more
important role. 

\begin{figure}[h]
\begin{center}
 \resizebox{8.4cm}{!}{
  \includegraphics{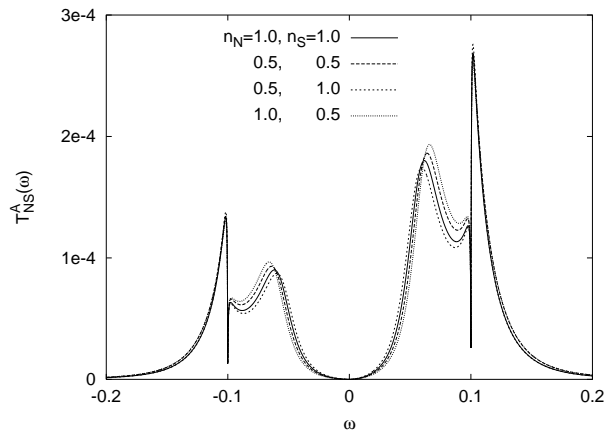}}
\end{center}
\caption{Andreev transmittance for various electron concentrations in the 
         leads. Model parameters are the same as in Fig. \ref{Fig10}.}
\label{Fig12}
\end{figure}

Now let's turn to the Andreev reflections and their modifications due to the
$e-h$ asymmetry. The Andreev transmittance $T^A_{NS}(\omega)$ (Eq. (\ref{GNS})),
shown in the Fig. \ref{Fig12} is also affected by the concentration of the
electrons in the leads. 
However quantitative behavior of $T^A_{NS}(\omega)$ seems to not depend on 
the $e-h$ asymmetry. The most pronounced qualitative differences occur for the 
energies $\omega = |E_d|$. Decreasing the number of electrons in the normal 
lead, $T^A_{NS}(\omega)$ also decreases around these energies. On the other 
hand, the effect is just opposite if we decrease the electron concentration in 
the $SC$ lead. One can rather easily explain the dependence of the 
$T^A_{NS}(\omega)$ on the number of normal electrons, taking note of the fact 
that probability of Andreev reflections is larger when number of electrons in 
$N$ lead is large and the number of holes small. However $T^A_{NS}(\omega)$ 
also depends on concentration of electrons and holes in the $SC$ lead. This can
be understood as follow: the number of electron-like (hole-like) quasiparticles 
in $SC$ is proportional to the concentration of the electrons (holes) in this
lead in the normal state. In the Andreev process, if two electrons enter the 
$SC$, the electron-like quasiparticle is created. This means that probability 
of Andreev reflections depends on the number of electron-like and hole-like 
quasiparticles in $SC$ lead. So increasing the number of holes in $SC$, the
probability of Andreev reflections of the impinging electrons is larger. This 
is exactly what we can read from the Fig. \ref{Fig12}. 

The modifications of the $T^A_{NS}(\omega)$ due to $e-h$ asymmetry are not so 
large as the modifications due to the asymmetry in the couplings 
(see Fig. \ref{Fig8}), nevertheless $e-h$ asymmetry influences the Andreev
tunneling.


\section{\label{sec7} Conclusions}

In conclusion we have studied a strongly interacting quantum dot connected to
the normal and superconducting leads. Using the equation of motion technique 
for the nonequilibrium Green's functions, we derived the formula for the 
current in terms of various tunneling processes. This technique allowed us to
calculate at once all the Green's functions emerging in the problem and 
perform consistent decoupling procedure for the higher order Green's functions.

We discussed the problem of the interplay between Kondo effect and Andreev
reflections. While the Kondo resonance is present in density of states, there 
is no zero bias anomaly in the differential conductance. As a matter of fact, 
the Andreev conductance is strongly suppressed for zero-bias voltages. We also 
further raised a question regarding the participation of the superconducting
electrons in the Kondo effect. The obtained results seem to support the 
scenario in which they do not participate in the Kondo effect. 

Finally, we discussed the problem of asymmetry in the couplings to the leads 
and found the large modifications of the Andreev conductance due to this 
effect, mainly for energies around dot level and superconducting gap. We also
studied the properties of the system when the concentration of electrons in the
leads can be changed. However, the modifications of the Andreev tunneling due 
to this effect are much smaller and quantitative only.



\end{document}